\documentclass[3p,times,twocolumn]{elsarticle}

\usepackage{ecrc}
\usepackage{txfonts}
\usepackage{natbib}
\usepackage{graphicx}

\volume{00}

\firstpage{1}

\journalname{Nuclear Physics B Proceedings Supplement}

\runauth{}

\jid{nuphbp}

\jnltitlelogo{{\small Nuclear Physics B Proceedings Supplement}}

\usepackage{amssymb}

\usepackage[figuresright]{rotating}

 \newcommand{\ssL}{{\scriptscriptstyle{L}}}
 \newcommand{\ssN}{{\scriptscriptstyle{N}}}
 \newcommand{\ssO}{{\scriptscriptstyle{O}}}  

\begin{document}

\begin{frontmatter}



\dochead{}


\title{%
NNLO massive corrections to Bhabha scattering and \\theoretical precision of
BabaYaga@NLO\tnoteref{t1}}
\tnotetext[t1]{Work supported in part by the Initiative and Networking Fund of the German Helmholtz Association``Physics at the Terascale", contract HA-101, 
by Sonderforschungsbereich/Trans\-re\-gio SFB/TRR 9 of DFG ``Com\-pu\-ter\-ge\-st\"utz\-te Theoretische Teil\-chen\-phy\-sik", 
by European Initial Training Network LHCPHENOnet PITN-GA-2010-264564 
and by Polish Ministry of Science and Higher Education 
from budget for science for 2010-2013 under grant number N N202 102638.}

 \author[label1]{C. M. Carloni Calame}
 \author[label2]{H. Czy\.z}
 \author[label2]{J. Gluza}
 \author[label2]{M. Gunia}
 \author[label3]{G. Montagna}
\author[label4]{O. Nicrosini}
\author[label4]{F. Piccinini}
\author[label5]{T. Riemann}
\author[label6]{M.~Worek}
 \address[label1]{School of Physics and Astronomy, University of Southampton, Southampton SO17 1BJ, U.K.}
 \address[label2]{Department of Field Theory and Particle Physics, Institute of Physics,
        University of Silesia, Uniwersytecka 4, PL-40-007 Katowice, Poland}
\address[label3]{Dipartimento di Fisica Nucleare e Teorica, Universit\`a di Pavia, and INFN, Sezione di Pavia, Via A. Bassi 6, 27100 Pavia, Italy}
\address[label4]{INFN, Sezione di Pavia, Via A. Bassi 6, 27100 Pavia, Italy}
\address[label5]{Deutsches Elektronen-Synchrotron, DESY, Platanenallee 6, 15738 Zeuthen, Germany}
\address[label6]{Fachbereich C Physik, Bergische Universit\"{a}t Wuppertal, Gaussstr. 20, D-42097 Wuppertal, Germany}

\begin{abstract}
We provide an exact calculation of next-to-next-to-leading order (NNLO) massive corrections to Bhabha scattering in QED, relevant for precision
luminosity monitoring at meson factories. Using  realistic reference event selections, exact numerical results for leptonic and hadronic corrections are given and 
compared with the corresponding approximate predictions of the event generator BabaYaga@NLO. It is shown that the NNLO massive corrections 
are necessary for luminosity measurements with per mille precision. At the same time they are found to be well accounted for in the generator 
at an accuracy level below the one per mille. An update of the total theoretical precision of BabaYaga@NLO is presented and possible directions for
a further error reduction are sketched.
\end{abstract}

\end{frontmatter}

\section{INTRODUCTION}
\label{intro}

Because of its experimental and theoretical characteristics, Bhabha scattering is the 
prime process used at $e^+e^-$ colliders to monitor their luminosity~\cite{Actis:2010gg}. At the GeV-scale 
$e^+ e^-$ colliders, from $\phi$ to $B$ factories, large angle Bhabha scattering is measured 
with experimental uncertainties at the per mille level. Correspondingly, the Monte Carlo 
(MC) generators used in the data analysis, like 
BabaYaga@NLO \cite{CarloniCalame:2000pz,Balossini:2006wc,Balossini:2006sd}, BHWIDE \cite{Bhwide1997} and 
MCGPJ~\cite{Mcgpj2006}, are precision tools that include the large logarithmic contributions
due to soft and collinear multiple photon emission matched with the exact next-to-leading 
order (NLO)
photonic corrections. In addition to the these ingredients, the MC codes include the effect of leptonic and 
hadronic vacuum polarisation, the latter computed using a data based routine for the evaluation of the 
non-perturbative light quark contribution \cite{Balossini:2008ht}. 

The above framework implies that the MC programs are affected by a theoretical uncertainty stemming 
mainly from missing (subleading) contributions at the level of NNLO radiative corrections. Therefore exact NNLO 
calculations are the benchmark for a reliable assessment of the error associated to the MC predictions.
Since the whole class of the NNLO QED corrections to Bhabha scattering became recently available 
(see \cite{Actis:2010gg} for a review), various
comparisons between the exact calculations and the MC results allowed to conclude that the theoretical
accuracy of the most precise luminosity tools is at the level of one per mille. Although this uncertainty seems
sufficient by comparison with the present experimental error, there is still room for a more reliable assessment 
and, presumably, a further error reduction. 

Here we focus on the NNLO massive corrections to Bhabha scattering, for which only preliminary 
results limited to leptonic corrections for a few experimental selections were used in the most 
recent, official estimate of the MC precision~\cite{Actis:2010gg}. These limitations are overcome in the present 
study \cite{CarloniCalame:2011xx}, wherein the full set of NNLO leptonic and hadronic corrections 
is computed and
all the event selections of experimental interest at meson factories are addressed. The exact
NNLO results are compared with the corresponding $O(\alpha^2)$ truncated predictions of the MC 
BabaYaga@NLO.

\section{THE NNLO LEPTONIC AND HADRONIC CORRECTIONS}

The full treatment of $N_f = 1,2$ massive corrections to Bhabha scattering requires the 
calculation of {\it i)} pure two-loop diagrams, including irreducible vertex and box 
corrections that are known to give rise to large leading $L^3, L = \ln(s/m_f^2)$, 
collinear contributions {\it ii)} loop-by-loop corrections  {\it iii)} real photon corrections
and  {\it iv)} corrections due to real lepton and hadron pair emission. The latter 
generate $L^3$ contributions canceling those from irreducible virtual loops. 

\subsection{Exact calculation}

Exact electron loop corrections ($N_f = 1$) were first computed in 
\cite{Bonciani:2004qt,Actis:2007gi}, while heavy fermion and 
hadronic loop contributions ($N_f  = 2$) were later calculated 
in \cite{Becher:2007cu,Bonciani:2007eh,Actis:2007fs,Actis:2008tr,Kuhn:2008zs}. As said above, these
virtual contributions must be combined with the real corrections. The cross section 
with complete NNLO leptonic and hadronic corrections can be written as
\begin{eqnarray}
\label{all-fermions}
 \frac{d \sigma^{\rm{\ssN\ssN\ssL\ssO}}_{N_f =1,2}} {d\Omega}
&=&
\frac{d \sigma^{\rm{\ssN\ssN\ssL\ssO}}_{\rm virt}} {d\Omega} 
+
\frac{d \sigma^{\rm{\ssN\ssL\ssO}}_{\gamma,\mathrm{soft}}(\omega)} {d\Omega} \nonumber\\
&+&
\frac{d \sigma^{\rm{\ssN\ssL\ssO}}_{\gamma,\mathrm{hard}}(\omega)} {d\Omega} 
+
\frac{d \sigma^{\rm{\ssL\ssO}}_{\rm real}} {d\Omega} 
\end{eqnarray}
which can be rearranged as 
\begin{eqnarray}
 \frac{d \sigma^{\rm{\ssN\ssN\ssL\ssO}}_{N_f =1,2}} {d\Omega}
\equiv
\frac{d \sigma^{\rm{\ssN\ssN\ssL\ssO}}_{\rm v+s} (\omega)} {d\Omega}
+
\frac{d \sigma^{\rm{\ssN\ssL\ssO}}_{\gamma,\mathrm{hard}}(\omega)} {d\Omega} 
+
\frac{d \sigma^{\rm{\ssL\ssO}}_{\rm real}} {d\Omega} 
\label{eq:NNLO}
\end{eqnarray}
where $\omega$ is a soft-hard photon separator. The terms entering 
Eq.~(\ref{eq:NNLO}) have been computed using various semi-analytical and MC
tools developed over the years, {\it i.e.} by the package bha\_nnlo\_hf  \cite{ACGR-bha-nnlo-ho:2011} for the 
virtual + soft corrections $d \sigma^{\rm{\ssN\ssN\ssL\ssO}}_{\rm v+s} (\omega) / d\Omega$, the MC 
code BHAGEN-1PH  \cite{chunp,Caffo:1996mi}
for hard photon radiation $d \sigma^{\rm{\ssN\ssL\ssO}}_{\gamma,\mathrm{hard}} (\omega) / d\Omega$,
and the generators HELAC-PHEGAS  \cite{Cafarella:2007pc} and 
EKHARA  \cite{Czyz:2006dm,Czyz:2010sp} for real lepton and hadron pair
emission $d \sigma^{\rm{\ssL\ssO}}_{\rm real} / d\Omega$, respectively.
For the $R$ contribution to the hadronic vacuum polarisation we adopt the 
recent compilation encoded in the routine VP\_HLMNT\_v2\_0~\cite{teubner:2010,Teubner:2009zz,Hagiwara:2006jt}.
Strictly speaking, EKHARA allows the computation of real pion pair emission only 
and no other generator exists for the calculation of similar processes containing other hadron pairs.
However, we observed in our numerical study~\cite{CarloniCalame:2011xx} that the cross section for real 
pion pair emission is always very small at all meson factories (at the level of $10^{-5}$ 
times the Born cross section or below it). Therefore an educated guess is that the processes of 
pair emission containing hadrons heavier than pions have a negligible effect in
our numerical predictions.

\subsection{Exact numerical results}

In Table \ref{table-NNLO-net} we show the 
relative contribution of the exact NNLO massive corrections for all the interesting experiments
at meson factories and using realistic reference event selections used 
in luminosity measurements. In Table \ref{table-NNLO-net} $S_{x}= \sigma_{x}^{\rm{\ssN\ssN\ssL\ssO}} / \sigma_{\rm BY}$ with 
$x=e^+e^-,  lep, had, tot$, where $tot$ stands for the sum of leptonic ($lep$) 
and hadronic ($had$) corrections, 
 and $\sigma_{\rm BY}$ is the full cross section from BabaYaga@NLO. We refer 
 to  \cite{CarloniCalame:2011xx} for details about the cuts used in our analysis.
 
 Some comments are in order here. 

\begin{table}[hbt]
\setlength{\tabcolsep}{0.2pc}
\caption{Exact relative NNLO massive corrections $S_{x}= \sigma_{x}^{\rm{\ssN\ssN\ssL\ssO}} / \sigma_{\rm BY}$, 
in per mille.}
\label{table-NNLO-net}
\begin{center}
\begin{tabular}{lclccc}
\hline
 & $\sqrt{s}$ & $S_{e^+e^-}$ &
 $S_{lep}$ &$S_{had}$ &
$S_{tot}$
\\
\hline
KLOE   & 1.020 & -3.935(5) &-4.472(5)  &1.02(4)    & -3.45(4)    \\
BES    & 3.097 & -2.246(8) &-2.771(8)  &--          &--           \\
BES    & 3.650 & -1.469(9) &-1.913(9)  &-1.3(1)    & -3.2(1)     \\
BES    & 3.686 & -1.435(8) &-1.873(8)  &--          &--           \\
BaBar  & 10.56 & -1.48(2)  &-2.17(2)   &-1.69(8)   & -3.86(8)    \\
Belle  & 10.58 & -4.93(2)  &-6.84(2)   &-4.1(1)    & -10.9(1)    \\
\hline
\end{tabular}
\end{center}
\end{table}

\begin{table*}[hbt]
\setlength{\tabcolsep}{.8pc}
\newlength{\digitwidth} \settowidth{\digitwidth}{\rm 0}
\catcode`?=\active \def?{\kern\digitwidth}
\caption{Comparison of the exact massive NNLO with BabaYaga@NLO results,
in per mille.}
\label{table-compar-NNLO-net}
\begin{tabular*}{\textwidth}{lclccccc}
\hline
 & $\sqrt{s}$ &  & $\sigma_{\rm BY}$ (nb) & $S_{e^+e^-}$ &
 $S_{lep}$ &$S_{had}$ &
 $S_{tot}$
\\
\hline
KLOE   & 1.020 &  NNLO   &     &-3.935(5) &-4.472(5)  &1.02(4)    & -3.45(4)  \\
       &       & BabaYaga &   455.71        &-3.445(2) &-4.001(2)  &0.876(5)   & -3.126(5)  \\
BES    & 3.097 &  NNLO    &     &-2.246(8) &-2.771(8)  &--          &--           \\
       &       & BabaYaga &  158.23         &-2.019(3) &-2.548(3)  &--          &--           \\
BES    & 3.650 &  NNLO    &     &-1.469(9) &-1.913(9)  &--1.3(1)    & -3.2(1)     \\
       &       & BabaYaga &  116.41          &-1.521(4) &-1.971(4)  &-1.071(4)  & -3.042(5)  \\
BES    & 3.686 &   NNLO  &     &-1.435(8) &-1.873(8)  &--          &--           \\
       &       & BabaYaga &  114.27          &-1.502(4) &-1.947(4)  &--         &--           \\
BaBar  & 10.56 &  NNLO   &      &-1.48(2)  &-2.17(2)   &-1.69(8)   & -3.86(8)  \\
       &       & BabaYaga &   5.195         &-1.40(1)  &-2.09(1)   &-1.49(1)   & -3.58(2)   \\
Belle  & 10.58 &  NNLO   &      &-4.93(2)  &-6.84(2)   &-4.1(1)    & -10.9(1)   \\
       &       & BabaYaga &  5.501          &-4.42(1)  &-6.38(1)   &-3.86(1)   & -10.24(2)   \\
\hline
\end{tabular*}
\end{table*}

According to the above discussion, the results for
the hadronic corrections have been obtained using the full $R$ parameterization for 
the virtual corrections and the contribution of pions only for real pair emission. Moreover, 
we do not provide results for the hadronic corrections at the BES (center of mass) c.m. energies 
on top of the $J/\psi$ and $\psi(2S)$ resonances. Actually, we observed 
\cite{CarloniCalame:2011xx} that for such 
energies the NNLO Bhabha cross section is dominated by the contribution of 
narrow resonances that can not be treated like mere perturbative effects and require a 
separate study.

Among the corrections induced by the different particle species, it can be seen from 
 Table \ref{table-NNLO-net} that the dominant contribution is given by the 
electron pairs,  with an increasing importance of muons and hadrons at the $B$ factories, 
the $\tau$ contribution being always negligible at meson factories.

In particular, among the real emission processes we concluded that only the reaction
$e^+ e^- \to e^+ e^- e^+ e^-$ gives significant contributions to the cross section used in the luminosity
measurements. When the accuracy of the experiment reaches the per mille level,  this process has to be
considered and its contributions added to the theoretical cross section or alternatively subtracted as a
background from the experimental cross section.

The main conclusion of the exact NNLO calculation is that the total correction is about
0.3 -- 0.4\% at KLOE, BES and BaBar and reaches about 1\% at Belle. 
Therefore, complete NNLO massive corrections, or a their approximation, are certainly needed 
for precise luminosity calculations.

\subsection{The NNLO massive corrections in the code BabaYaga@NLO}

BabaYaga@NLO is one of the most precise and widely used 
theoretical tools to monitor the luminosity of meson factories. It is a QED Parton Shower 
generator, whose intrinsic leading logarithmic accuracy is improved by the 
inclusion of exact NLO soft+virtual and hard photon correction factors.
In BabaYaga@NLO these NLO factors are dressed by self-energy insertions, 
so that a subset of the complete NNLO massive corrections 
previosuly discussed is included in the code.
More precisely, the contributions accounted for are: {\it i)} 
the factorizable loop-by-loop corrections within the full class of NNLO virtual corrections
{\it ii)} the contribution of real (soft and hard) photon emission in the sector 
of real corrections. In other words, in the generator any contribution from real pair radiation 
and irreducible virtual corrections is neglected, which makes the approach theoretically consistent 
because it avoids an imbalance of $L^3$ contributions.

Both leptonic and hadronic self-energy contributions are taken into account in the code. 
For the $R$ parameterization we use in the present study the same VP\_HLMNT\_v2\_0
routine as in the exact calculation, in order to perform consistent comparisons.

\section{EXACT NNLO MASSIVE CORRECTIONS VS. BABAYAGA@NLO}

The quality of the approximation inherent in BabaYaga@NLO is shown in Table 
\ref{table-compar-NNLO-net} in comparison with 
the results of the exact calculation. Again realistic reference event selection cuts are considered.

It can be seen from Table 
\ref{table-compar-NNLO-net} 
that there is a generally good agreement for all the meson factories experiments, 
the relative differences always being below the one per mille. There is a maximum registered
difference of about 0.07\% at Belle, where the corrections have the largest impact, but the 
agreement is at the level of a few units in $10^{-4}$ for all the other experiments. Therefore
the very bulk of NNLO massive corrections is accounted for in BabaYaga@NLO.

\begin{figure}
\caption{The relative difference, in per mille, of the NNLO massive leptonic and hadronic corrections
between exact $\sigma_{exact}^{NNLO}$ and BABAYAGA@NLO $\sigma_{BY}^{NNLO}$, as a function of 
an acollinearity cut for two different angular
acceptance regions for KLOE-like event selections.}
\includegraphics[width=0.5\textwidth]{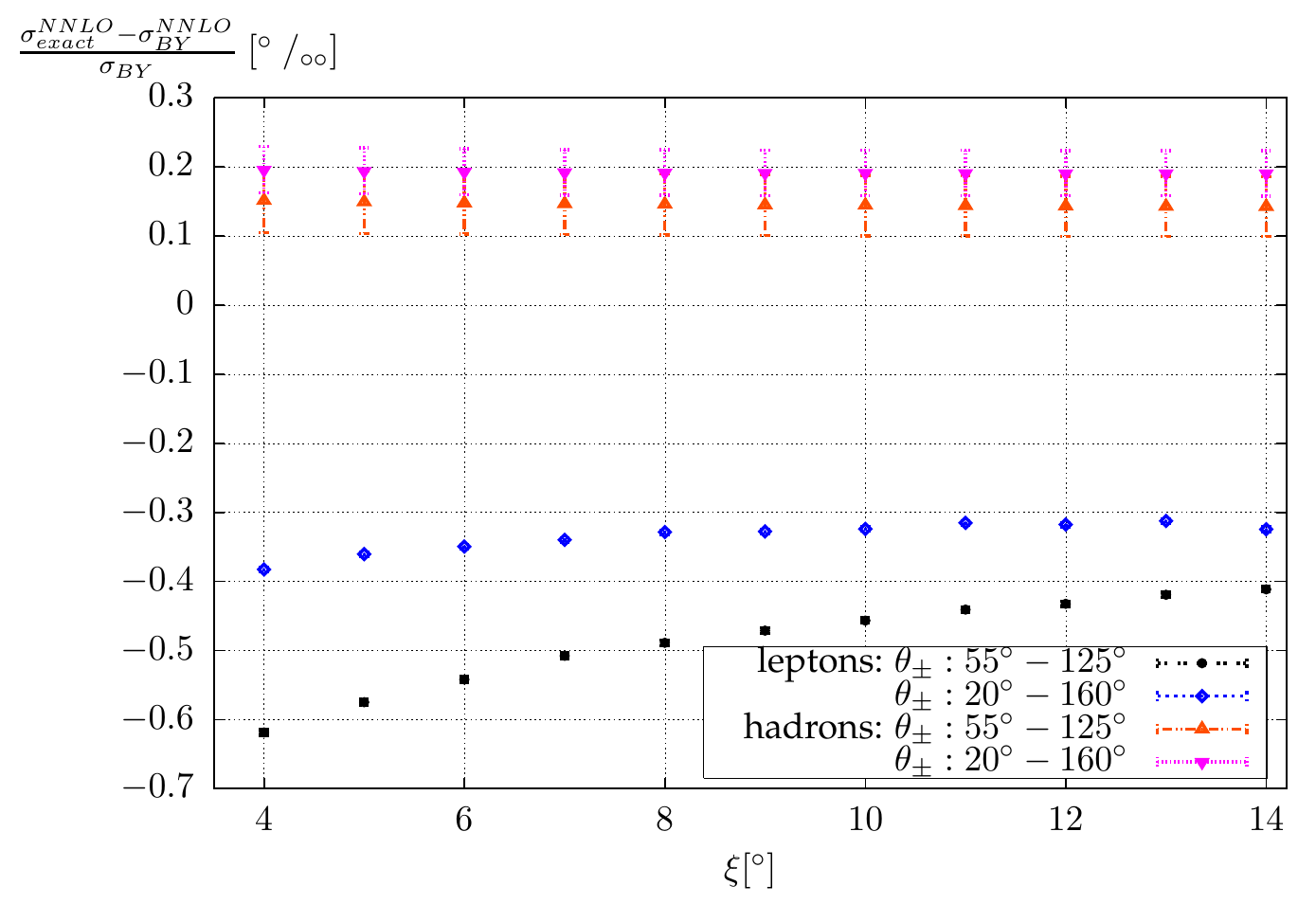}
\label{fig:kloe}
\end{figure}

In our analysis, we also studied how the above conclusions are stable against variation of the 
experimental cuts, such as acceptance and acollinearity cuts. We noticed that for 
the leptonic corrections there is a slight increase in the difference for particularly
tight acollinearity cuts but in general the difference is not particularly 
sensitive to the variation of the event selection. In any case, also for particularly severe 
cuts, the maximum observed difference still remains 
below the one per mille~\cite{CarloniCalame:2011xx}. An example of such a cut-dependence 
study is shown in Fig. \ref{fig:kloe} for KLOE, which is the experiment with the smallest uncertainty  
in the measurement of the luminosity, at the level of 0.2\%. It can be seen that a variation of
the acollinearity cut from its reference value of $9^\circ$ to smaller values only slightly 
affects the relative difference between the exact calculation and the BabaYaga@NLO approximation
for leptonic correction, leaving the situation for hadrons practically unchanged.

\section{CONCLUSIONS}

We provided an exact calculation of the NNLO massive corrections to Bhabha scattering in QED,
exploiting a number of analytical results and computational tools developed over the years.
We presented exact numerical predictions for leptonic and hadronic corrections in the
presence of realistic cuts for luminosity monitoring at meson factories. We compared the 
exact results with the approximate ones provided by the MC code BabaYaga@NLO. 
We concluded that NNLO massive corrections are well accounted for in the generator 
at an accuracy level below the one per mille. This reinforces the estimate of the total theoretical precision of BabaYaga@NLO, 
previously established on the grounds of only partial results for NNLO massive corrections. 
The present total accuracy of BabaYaga@NLO can be estimated 0.1\% at KLOE and
BES (excluding data taking on top of narrow resonances $J/\psi$ and $\psi(2S)$), and about 0.15\% 
at the $B$ factories.

This precision is today sufficient by comparison with the experimental accuracy
of meson factories. Nonetheless, there is still room for a further error reduction along the
following directions. There is a need for an assessment of the theoretical accuracy for 
luminosity measurements in a close vicinity of narrow resonances. This requires detailed studies
of the uncertainties associated to the hadronic vacuum polarisation and NNLO hadronic
corrections, including beam spread effects. 
The leading part of the presently missing NNLO massive corrections in BabaYaga@NLO
could be implemented through {\it e.g.} QED Structure Functions, to account for the interplay
between virtual irreducible corrections and lepton pair radiation. Last but not least, the recently 
obtained exact results for the one-loop corrections to radiative Bhabha scattering 
\cite{Actis:2009uq}, as well as 
for $e^+ e^- \to \mu^+ \mu^- \gamma$ \cite{Actis:2009uq,ksy:2099app}, 
are the final
NNLO benchmark that should be deeper investigated in comparison with the 
present MC approximation.

All these perspectives are left to future works.

\vskip 12pt\noindent
\leftline{\bf Acknowledgments}

We would like to thank Thomas Teubner for useful discussions on vacuum polarisation and Wang Ping 
for providing us some of the reference luminosity event selections used in our study. 

The collaboration
between the authors of the present work originated during the activity of 
the ``Working Group on Radiative Corrections and Monte Carlo Generators for Low Energies" 
[http://www.lnf.infn.it/wg/sighad/]. 

One of us (G. M.) wishes to thank Simon Eidelman for the invitation to PHIPSI11, and is grateful to 
all the
organizers for the warm hospitality and the stimulating atmosphere of the workshop.








\end{document}